\documentclass[aps,amsmath,prl,twocolumn,superscriptaddress,nofootinbib]{revtex4-2}
\usepackage{graphicx}
\usepackage{bm}
\usepackage{epsfig}
\usepackage{amsmath,latexsym}

\newcommand{\nn}{\nonumber}

\newcommand{\beq}{\begin{equation}}
\newcommand{\eeq}{\end{equation}}
\newcommand{\bea}{\begin{eqnarray}}
\newcommand{\eea}{\end{eqnarray}}

\begin{document}

\title{First Principles Prediction of the Landau Parameter  for  Fermi  Liquids  near the Unitarity Limit}
\author{$^{1}$Shashin Pavaskar and $^{1}$Ira Z. Rothstein \vspace{0.7 cm} \\ $^{1}$ Department of Physics, Carnegie Mellon University, Pittsburgh, Pennsylvania, 15213, USA}


\begin{abstract}
This paper explores the  behavior of systems of cold fermions  as they approach  unitarity  above the critical temperature.
 As we move away from unitarity, by decreasing the scattering length, 
the dilaton, the Goldstone boson resulting from the spontaneous breaking of Schrodinger symmetry by the Fermi sea, becomes gapped.
At  energies below this gap, the interaction between quasi-particles will be dominated by local interactions generated by off-shell 
dilaton exchange. The dilaton mass can, in turn, be related via anomaly matching, to the scattering length and contact parameter
within the confines of a systematic expansion. 
We use this relation to  predict  the s-wave  Landau parameter to be
 $f=\frac{4\pi a (2\epsilon(p_F)-p_F^2/m_\star)^2  m}{3p_F^4 \tilde{C}(a)}$
where $a$ is the scattering length,  $m$ the atomic mass, $m_\star$, the effective mass which can be extracted from heat capacity,  and  $\tilde {C}(a)$ is the dimensionless contact parameter.  The range of validity of this prediction  (given in eq.(\ref{bound})) is determined by the value of contact parameter and Fermi velocity, which depend upon the scattering length.
It is expected to be valid in  a range  above   $k_F a \sim 1$, but the actual window will depend upon the values of aforementioned parameters.
 Given this result  for $f$, we predict the   compressibility, spin susceptibility  and the quasi-particle life-time.  

\end{abstract}

\maketitle
\section{Introduction}

The low energy description of Fermi liquids has been textbook material for many years now. Above the critical temperature one can make
predictions for  observables  in terms of a collection of material dependent  Landau parameter(s) $f_l$, to leading
order in an expansion in $E/E_F$. Whether considering a metal, or a gas, in general, these couplings are  treated as unknown
parameters whose values can be determined by independent measurements.
However, one might hope that for systems with enhanced symmetries, the couplings might be predictable. 
As such, it would seem that fermions at unitarity would be a compelling system, as it manifests the
maximal  Schrodinger group symmetry. These systems exhibit universal behavior as a consequence of
the divergent scattering length.  Recent experimental progress in producing such ``uniform quantum gases" \cite{exp} via boxed traps, has opened the door to the study of such ideal systems.

However, fermions in the unitary limit can not be described by the canonical Fermi liquid EFT (as described e.g. in \cite{eft})  because there is no way to non-linearly realize the   spontaneously broken conformal and
boost invariance and   maintain Fermi liquid behavior, as shown in \cite{RS,RS2}.
At present we do not know how to calculate in a systematic expansion in the unitary limit. Here we will instead  calculate
far enough away from  unitarity  that we can treat it as a Fermi liquid but close enough to keep some approximate
symmetries. By doing so we will able to predict the aforementioned Landau parameters in a regime where
weak coupling $k_F a$ calculations fail. Making any systematic first principle prediction for a strongly coupled
theory is an extreme theoretical challenge, and we manage to do so only in a narrow range of parameter space.

To understand how to calculate near unitarity we must first ask why Fermi liquid theory breaks down at unitarity
where the atomic underlying theory is invariant under the full non-relativistic conformal (Schrodinger) group.  The existence of the Fermi sea breaks a subset of  symmetries: three boosts, dilatations and special conformal transformations. While the breaking of global internal  symmetries leads to gapless Goldstone modes, one per broken 
generator,   when spacetime symmetries are broken, this is no longer true \cite{ivanov,volkov}. In such a case, the Ward identities can be saturated 
by excitations which can be arbitrarily wide, i.e. they need not be quasi-particles \footnote{For a recent discussion of this issue see \cite{alberte}.}. At the level of the action, invariance may 
be maintained despite the dearth of Goldstones. The modes   for which the corresponding broken generators' commutator with  unbroken translations yields another broken generator (not in the same multiplet)
can be eliminated from the action.  
This is called the Inverse Higgs mechanism (IHM) and one can use   the space-time coset construction \cite{ivanov,volkov} to determine invariant constraints which eliminate the extra Goldstones.
However, there are cases where there are no IHM's at play and yet the Goldstones, which seemingly should be in the spectrum, are not. The classic example of this is $He^3$ where only boosts 
are broken, and  there are no corresponding Goldstones.  In such systems, dubbed ``framids" \cite{zoo},  the symmetry is realized by constraining the form of the interactions \cite{RS2}.
In fact, the famous Landau conditions on Fermi liquids is the constraint that must be imposed on the action to ensure boost invariance. Such a condition can be   considered  a ``Dynamical Inverse Higgs Constraint" (DIHC) \cite{RS2}. In \cite{RS} it was shown that in the unitary limit, in three spatial dimensions, the symmetries can be realized by either imposing another DIHC or by the inclusion of a dilaton. In either case the
system cannot behave like a canonical Fermi liquid above $T_c$.




Fermi liquid theory starts with the assumption   that
 quasi-particles (in our case fermionic) exist in the spectrum  with widths that scale as $\Gamma \sim E^2$, due to Pauli blocking.
Such  systems will have two marginal couplings, the ``BCS'' and forward scattering channels, with the former
growing strong in the IR leading to breaking of the particle number $U(1)$ symmetry.
%
%
%
However,  when interactions in the UV become strong,  the Fermi liquid description can break down at which point there 
may no longer be any stable quasi-particles, leading to non-Fermi liquid behavior. Such is the case for fermion in the unitary limit.

In this paper we explore the approach to this non-Fermi liquid behavior by  calculating how the quasi-particle width begins as a function of  the scattering length ($a$). The starting point is the effective field theory of Fermi liquids \cite{eft} where we consider small fluctuations around the Fermi surface. We are interested in studying the normal phase of the theory where $T>T_c$.  
Furthermore, as will be explained below, to maintain calculational control we will keep the scattering length finite yet
large, where canonical perturbative methods fail \cite{hammer}.

 Our approach begins by utilizing the pattern of spontaneous breaking of space time symmetries.
    In \cite{RS}  it was shown that at unitarity, non-Fermi liquid  behavior emerges due to the presence of a non-derivatively coupled gapless Goldstone (the dilaton) that arises as a consequence of the symmetry breaking pattern. 
  Typically Goldstones are derivatively coupled and therefore decouple in the far IR, however,
for spontaneously broken space-time symmetries,   for certain symmetry breaking patterns,  Goldstone bosons, such as the dilaton, couple  non-derivatively \cite{vish,RS2} leading to a strong coupling  in the infra-red. 
 
When we perturb away from unitarity, the dilaton gets gapped,  with its mass  acting as a control parameter which can be used to study the cross-over behavior.
When the mass is non-vanishing but sufficiently small,  Fermi liquid behavior is expected and dilaton exchange will dominate the fermion-fermion interaction.
Moreover, the dilaton mass  can be determined by matching  the conformal anomaly, between the UV theory (where it is exactly known) and
the IR theory. Using this result, along with the fact that the dilaton coupling is fixed by symmetry, allows us to
 to predict the s-wave Landau parameter   in terms of the scattering length, the effective mass of the fermion  and the contact parameter.
With this result in hand we then predict the value of the  compressibility, spin susceptibility and the quasi-particle lifetime.

{\bf Notation and Units:} We will be working with $\hbar=m=1$ unless otherwise states, where $m$ is the bare quasiparticle mass.



\section{The EFT}

In the normal phase of a gas of cold atoms the only spontaneously broken symmetries are  boosts. 
Despite this fact, the spectrum has no Goldstone bosons and  the broken boosts are still non-linearly realized via the non-trivial  (Landau) relation  between the effective mass and the p-wave Landau parameter.

The unitary limit in the trivial vacuum is a point of enhanced symmetry realizing the full thirteen parameter Schrodinger group.
 The Fermi surface spontaneously breaks boosts (K), dilatations (D) and special conformal transformations(C). 
 The way  these broken symmetries can be realized was discussed in \cite{RS,RS2} which for completeness we summarize here.
In the case at hand, the Goldstone associated with the breaking of conformal symmetry can be eliminated using the IHC arising from the relation 
 \begin{equation}
     [H,C] = iD,
 \end{equation}
leaving only the dilaton, the Goldstone associated with the broken scale invariance. The boost Goldstone called the framon is necessary to write down a Galilean invariant action for the dilaton. However, it was shown in \cite{RS2}, that one can eliminate the framon using an operator constraint called the Dynamical Inverse Higgs constraint (DIHC). In the Fermi liquid theory, the DIHC is nothing but the aforementioned Landau relation. 
The logical possibility remains that the action obeys further constraints, such that there is no dilaton in the action.
However, as shown in \cite{RS2}, without a dilaton in the action the quasi-particle would have to obey a quadratic
dispersion relation  (as opposed to linear) and the coupling would have to undergo power law running. Moreover, independent of the choice of field variables, at unitarity there still must
be a cut in the stress-energy correlation function corresponding to a 
 highly damped excitation. 
Moving away from unitarity towards a quasi-particle description, this gapped channel will be nothing but the massive dilaton.

Let us explore the consequences of the existence of a  light ($m_\phi \ll E_f$) dilaton in the spectrum.
We will treat the  dilaton mass as the leading order perturbation in the conformal symmetry breaking, with higher order
corrections being down by powers of $m_\phi/E_F$. We begin by first writing down the action in the conformal/unitary  limit.
Since the scattering length diverges in this limit, the only scale in the theory is the Fermi energy $E_{F}$. To write down the action for quasi-particles and the dilaton, we utilize the technique of spacetime coset constructions \cite{ivanov,volkov} which is a systematic way of non-linearly realizing the symmetries. We present here the results given in \cite{RS} and refer the reader 
to  that paper for details. 

 At the unitary point, the coset element can be written as 
\begin{equation}
   U = e^{iHt}e^{-i\vec{P}.\vec{x}} e^{-i\vec{K}.\vec{\eta}}e^{-i D\phi}e^{-i C\xi}
\end{equation}
where $\vec{\eta}(x,t), \phi(x,t)$ and $\xi(x,t)$ are the framon, dilaton and Goldstone of the conformal transformation respectively. In the remainder of the work, we will explicitly drop any $x$ and $t$ dependence from the fields. Using the Maurer-Cartan (MC) form, one can extract the covariant derivatives for the Goldstones which transform linearly under the broken group. 
The  coupling of the dilaton $\phi$ in the quasi-particle action is given by 
\begin{eqnarray}
     S_{\psi}= \int dt  d^{3}x \:  {\psi_{\sigma}^{\dagger}}( i\partial_{t}{\psi}_{\sigma} - e^{\frac{2\phi}{\Lambda}}\epsilon(e^{\frac{\phi}{\Lambda}}i\vec{\partial}){\psi}_{\sigma}  ) +  \frac{f_0}{2} ( \psi^\dagger_{\sigma}  \psi_{\sigma})^2 
   ] , \nn \\
 \end{eqnarray} 
 We have kept only the $l=0$ Landau parameter. The addition of higher $l$'s will not change our predictions as we shall see.


 We have introduced a scale $\Lambda$ to normalize the dilaton field in the exponential. Under dilatations, the dilaton shifts by a constant $\phi \longrightarrow \phi + c\Lambda $ whereas the coordinates transform as $t \longrightarrow e^{2c}t$ and  $x \longrightarrow e^{c}x$. The quasi-particle fields and their covariant derivatives have to transform as a linear representation of the unbroken group $\psi (x,t) \rightarrow e^{-\frac{3}{2} c}\psi (x,t) $.
 One is free to add an invariant  term of the form $V_{dil} = C e^{-5\phi/\Lambda}$ to the dilaton Lagrangian. Thus maintaining a light dilaton implies $C$ must be fine tuned to be small, as its natural value is of order of the cut-off.
 This is analogous to the cosmological constant problem, the most egregious fine tuning in nature.
 However, in the context of fermions at unitarity, the appropriate fine tuning is achieved by choosing the magnetic
 field such that the atomic system is sitting near the Feshbach resonance. 

Expanding  around the Fermi surface to leading order in the dilaton field $\phi$ in the quasi-particle action, 
\begin{equation}
S_{\psi} = \int d^{3}x  \hspace{0.05cm} dt \hspace{0.1cm}   \frac{\phi}{\Lambda}\psi^{\dagger}_{\sigma} \psi_{\sigma} ( 2 \epsilon(k_F)- \vec \partial_{p}\epsilon(k_F) \cdot \vec k_F)+...
\end{equation}
where we have dropped terms sub-leading in the power expansion, since momenta normal to the Fermi surface scale as $\lambda$.
It is convenient to re-express this coupling in terms 
the Fermi velocity  $\vec v_F \equiv \vec \partial_{p}\epsilon(k_F)$ then
\begin{equation}
S_{\psi} = \int d^{3}x  \hspace{0.05cm} dt \hspace{0.1cm}   \frac{\phi}{\Lambda}\psi^{\dagger}_{\sigma} \psi_{\sigma} \left(  2 \epsilon(k_F)- v_F k_F \right)+...
\end{equation}
Notice  that in the free limit, $\epsilon(k_F)= k_F^2/2m$, this coupling vanishes.
For notational convenience we define  
\beq
\label{delta}
 \delta E\equiv \left(  2 \epsilon(k_F)- v_F k_F \right) 
\eeq
 which quantifies deviation from the canonical dispersion relation. For systems near unitarity $\delta E/E_F < 1$.

  Power counting dictates that the dilaton momenta must scale homogeneously under an RG transformation in all directions ($\vec{p} \rightarrow \lambda\vec{p}$)
  and thus will only scatter nearby points on the Fermi surface. Any other choice of scalings would lead to
  a power suppression.
   The quasi-particle and the dilaton energies scale in the same way as we move towards the Fermi surface ($\omega \sim  \lambda\omega$). From the kinetic terms in the dilaton and quasi-particle actions, we can read off the scaling of the momentum space dilaton and quasi-particle fields   
\begin{equation}
         \psi(p,t) \sim \lambda^{-1/2} \qquad \phi(p,t) \sim \lambda^{-2}.
\end{equation}
The   scaling of the dilaton-quasi-particle interaction is  marginal as can be seen by going to momentum space and
noting that, as in the four point quasi-particle interaction, the  delta function enforcing the three-momentum conservation  scales as $1/\lambda$ while the momentum space measure will scale as
\beq
d^3p_1 d^3p_2 d^3k \sim \lambda^5,
\eeq
as all three momentum components of the dilaton, as well as the quasiparticle momenta along the direction normal to
the Fermi surface,  scale as $\lambda$.

\section{The approach to Non-Fermi Liquid Behavior}
   As we move away from the unitary point, the scattering length becomes finite and scale invariance becomes an approximate symmetry of the effective theory. Hence the dilaton becomes a  gapped pseudo-goldstone.  As we will see, we can determine the mass of dilaton in terms of the scattering length and the contact parameter.  We are working in the units where the fermion mass is one and $\hbar=1$, the length dimensions will be 
\begin{equation}
[t] = 2 \qquad [\phi]=-\frac{1}{2}  \qquad [\psi]= -\frac{3}{2}.
\end {equation}

Away  from unitarity, 
the conformal symmetry is explicitly broken, however if we keep the scale of explicit symmetry breaking (the inverse scattering length)   small
compared to the scale of spontaneous symmetry breaking (the Fermi wave number) we may still treat the dilaton as a pseudo-Goldstone boson. 
The smallness of the dilaton mass  follows from the fact that the scattering length is tuned to be large.
The mass of the dilaton is treated as a spurion such that the action is invariant if we scale it according to its dimensions.
  \begin{equation}
      \delta L = \frac{1}{2}m_{\phi}^{2}\phi^{2}
  \end{equation}
We now use a matching procedure to calculate  $m_{\phi}$. In the effective theory away from unitarity, the scale current is not conserved.  


 \begin{equation}
     \partial_{\mu}s^{\mu} = m_{\phi}^{2}\Lambda \hspace{0.1cm}\phi 
 \end{equation}
 
 
 
 We will use current algebra to extract the mass by matching it onto the full theory result.  From the Noether construction the dilatation charge is given by
 \begin{equation}
     D^{0}(0) = \Lambda \int d^{3}x \hspace{0.1 cm}  \pi(\vec{x},0)
 \end{equation}
 where $\pi(x)$ is the conjugate momentum to $\phi$. Hence using (9) we have 
 \begin{equation}
 \label{split}
 \begin{split}
        \int_{x} \hspace{0.1cm} [D^{0}(0),\partial_{\mu}s^{\mu}(\vec{x},0)] 
        &=  \int d^{3}x \hspace{0.1cm} m_{\phi}^{2}\Lambda^{2} 
 \end{split}
 \end{equation}
We match this commutator to the full theory, which is a   microscopic description of the theory, in terms of fermions with action 
\begin{equation}
\label{act}
   S = \int dt \int d^{3}x \hspace{0.1cm}i \chi^{\dagger} \partial_{t} \chi +  \frac{1}{2}\chi^{\dagger} \nabla^{2} \chi + g(\mu) (\chi^{\dagger}\chi)^2
\end{equation}
  where $\chi$ is two-spinor. The Van der Waals scale($\Lambda_{VDW}$) provides the upper cutoff in the theory that suppresses higher dimensional operators.  
  In  the renormalized coupling can be written in terms of the scattering length as
  \cite{Braaten:2004rn}
 \begin{equation}
     g(\mu) = \frac{4\pi}{-\frac{2}{\pi}\mu + \frac{1}{a}}.
 \end{equation}

The four-fermion interaction defined in (\ref{act}) explicitly breaks scale invariance. One can verify that the dilatation charge, the divergence of the scale current and their commutators are given respectively by 

 \begin{eqnarray}
     D^{0}(0)  
     =  \int d^{3}x (\frac{3}{2}\chi^{\dagger}(\vec{x},0)\chi(\vec{x},0) + \chi^{\dagger}(\vec{x},0)\vec{x}\cdot \vec{\partial}\chi(\vec{x},0)) \nn \\
 \end{eqnarray}
  \begin{equation}
    \partial_{\mu}s^{\mu} = (g(\mu) + \beta(g))(\chi^{\dagger}\chi)^2
 \end{equation}
 \begin{eqnarray}
 \label{res}
 \hspace{0.1cm} 
 \int d^{3}x \hspace{0.1cm} m_{\phi}^{2}\Lambda^{2} 
 &=& 3\int d^{3}x \hspace{0.1cm} (g(\mu)+ \beta(g)) \hspace{0.1cm}(\chi^{\dagger}\chi)^2. \nn \\
 \end{eqnarray}
 
 Where in (\ref{res}) we have matched the commutators in the full and the effective theory using (\ref{split}). 
 Note that the RHS of (\ref{res}),  is an RG invariant,  and  the dilaton mass is  independent of the scale $\mu$. The coupling and the four-fermion operator both depend on the scale $\mu$ but the dependence cancels exactly in 
 (\ref{res}) to give a scale independent mass as required.
  Evaluating the beta function and taking the expectation value, we have 
  \begin{eqnarray}
    m_{\phi}^{2}\Lambda^{2}  &= &
      \frac{3}{4\pi a }\langle g^{2}  \hspace{0.1cm}   \chi_{\uparrow}^{\dagger}\chi_{\uparrow}   \chi_{\downarrow}^{\dagger}\chi_{\downarrow} \rangle
     \equiv \frac{3}{4\pi a} \hspace{0.1cm} {\cal C}(a)
 \end{eqnarray}
 where we have now made the spin state explicit and  ${\cal C}(a)$ is the contact density \cite{Tan}  whose vacuum expectation value is  a measure of the local pair density of the fermions and is  independent of the RG scale $\mu$. For any system consisting of fermions with two spin states and large scattering length, one can define universal relations which depend on the contact. 
Note that $\Lambda$ is still an undetermined free  parameter.  However, we will see that it will cancel in the  calculation of the Landau parameter $f_0$.

 If the dilaton mass is sufficiently small it will dominate the quasi-particles interactions, as other contributions  to the  interaction, arising from integrating out other modes, will be parametrically  suppressed by powers of $m_\phi/E_F$, where $E_F=\frac{k_F^2}{2m}$.
 We integrate out the dilaton to generate  net interaction
\beq
L_{int}= \frac{1}{2}\Big(f_0 + \frac{4\pi a \:  \delta E^2}{3 \tilde{\cal{C}}(a) k_F^4}\Big)(\psi^\dagger_{\sigma}  \psi_{\sigma} ) (\psi_{\sigma'}^\dagger \psi_{\sigma'})
\eeq
 $\delta E$ is defined in eq.(\ref{delta}).
The corrections to this expression are suppressed by  powers of $ (E^2,T^2)_{max}/m_\phi^2 =\frac{ \pi}{3} k_F a \frac{ (E^2,T^2)_{max}}{\tilde{\cal{C}}(a) E_F^2}$ and we have
taken $\Lambda \sim k_F^{1/2}$ as the symmetry breaking scale.
 $\tilde {\cal C}(a)=\frac{{\cal C}(a)}{k_F^4}$ is the dimensionless contact parameter \cite{meas}.
 
Thus if we are in the regime
\beq
\label{bound}
\tilde{\cal{C}}(a) \left( \frac{E_F}{(E,T)_{max}}\right)^2  \gg a k_F\gg \: \tilde{\cal{C}}(a) \left( \frac{E_F}{\delta E}\right)^2  k_F f_0
\eeq
then the dilaton exchange dominates so that we have an effective coupling
 \beq
\label{f}
 f_D=\frac{4\pi a \: \delta E^2 \: m }{3  \tilde{{\cal{C}}}(a)  p_F^4},
\eeq
where we have re-introduced the factors of $\hbar$ and the atomic mass $m$.
It is important to keep in mind that $\delta E$ as defined in (\ref{delta})  depends upon $a$ itself 
in non-perturbative way that  can be determined from experiment or possibly simulations.
We will make some estimates for this dependence below to get a handle on the region of
validity of the prediction.

 Note that since the coupling to the dilaton is scalar in nature,   higher angular momentum interactions will be sub-leading
 in our expansion.  Since we are working in the unbroken phase $T_{min}=T_c$, where $T_c$ is the superfluid transition temperature.

 \section{Compressibility and Spin Susceptibility}
 
 We can now make a prediction for the compressibility ($\kappa$) in terms of $f_D$, again, as this contribution dominates in the strongly interacting region.  We may extract $\kappa$ from a canonical calculation in the effective theory of the number density response function and it  is given by
\begin{equation} 
\label{C} 
\kappa= \frac{\kappa_0}{1- \hbar^2 N_F f_D} =  \frac{\kappa_0}{1- \frac{1}{3\pi} \frac{m^{\star}}{m} \frac{k_F a \: \delta E^2  }{ \tilde{C}(a) E_F^2}} 
\end{equation}
where $N_{F} = \frac{m^\star p_F}{\hbar^3 \pi^2} $ is the density of states at the Fermi surface and $\kappa_0$ is the compressibility of the free Fermi gas. One can obtain $m^{\star}$ from the measurement of the specific heat of the Fermi gas  via the relation $C_v=\frac{m^{\star} p_F}{3 \hbar^3}k_B^2 T$.   
 From (\ref{C}) we find an additional limit of our EFT if we assume the compressibility  must be positive
 \beq \label{compressibilitybound}
ak_F<   3\pi\left( \frac{E_F}{\delta E}\right) ^2  \frac{m }{m^{\star}} \tilde{\cal{C}}(a).
 \eeq
Since the compressibility is a zero frequency observable, this is the appropriate bound on the validity of this particular prediction, 
as opposed to the upper bound stemming from the LHS\footnote{ Note that there is no thermal mass for the dilaton, due to quasi-particle loops, in the large $T$ limit \cite{chubakov}, though the width will scale with $\frac{\delta E}{\Lambda}T$. } of (\ref{bound}).

One can also consider the response of the Fermi liquid to an external magnetic field and calculate the spin susceptibility $\chi$. This can be computed straightforwardly in the effective theory and is given by 
\bea
\chi =  \frac{\chi_0}{1 + \hbar^2 N_F f_D} = \frac{\chi_0}{1 + \frac{1}{3\pi} \frac{m^{\star}}{m} \frac{k_F a \: \delta E^2  }{ \tilde{C}(a) E_F^2}} 
\eea
where $\chi_0$ is the susceptibility of the free Fermi gas. Note that the validity bound in \eqref{compressibilitybound} need not hold for the above prediction. The regime of validity for the spin susceptibility can only be clarified from experiments.  
 
\section{Quasi-particle width}
 We may also calculate the quasi-particle width  using our result for $f_D$.
 The self-energy only gets contributions from the  forward scattering coupling, as the other marginal coupling (BCS) is restricted to back-to-back interactions.
The imaginary part of the self energy of a fermi-liquid due to $S$-wave interaction is $\Gamma_{FL}(E,T)  = f_0^{2}I(E,T)$, where  $I(E,T)=\frac{ m_\star^{3} }{4 (2\pi)^{3}} (E^2 +  (\pi kT) ^2) $ is the imaginary part of the two-loop self energy diagram at finite temperature\cite{chubakov}.  
Using our result (\ref{f}) we can then calculate the quasi-particle width 
\begin{equation}
\begin{split}
 \Gamma(E,T) &= f_{D}^{2}I(E,T). \\
\end{split}
\end{equation}
such that, 
\begin{equation}
\label{predict}
 \Gamma(E,T)= \frac{ \: m^{\star}}{18 \pi \: {\tilde {\cal C}(a)}^2 }\left(\frac{m^{\star}}{m}\frac{ a \: \delta E^2}{4\hbar E_F^2} \right)^2 (E^2 +  (\pi kT) ^2)    
\end{equation}
This prediction is valid in the range defined by eq.(\ref{bound}).

The  theoretical errors in this predictions are of order
\beq
\frac{\Delta \Gamma_{T}}{\Gamma}\sim O\left( \frac{1}{k_F a} \right)+ O\left(k_F a\left(\frac{E^2}{E_F^2}\right) \right).
\eeq


 



 \vskip-1in
\section{Conclusions}
It is known that degenerate fermionic systems cross over from Fermi to non-Fermi liquids as unitarity is approached.
Symmetry requires that  Fermi gases at unitarity manifest  a gapless excitation  in response to
external stress. This ``dilaton'' mode will look like an over damped sound mode, but it might be hoped to be  isolated since we are 
working in the attractive regime where there is no zero sound.  Furthermore, by working 
below the hydrodynamic limit, there will be no contamination from second sound. 

Perturbing away from the unitary limit gaps this mode. For energy scales large compared to the gap,
the quasiparticle excitations are expected to behave as in a non-Fermi liquid with a width that scales linearly
with the energy. However, as the energy of the quasi-particle drops below the gap the dilaton mediated
interaction localizes and Fermi liquid behavior with the width scaling quadratically with energy is expected.
The behavior of the system as a function of energy and scattering length is depicted in figure one.

The key insight noted here is that the mass of the dilaton can be fixed by matching the effective theory current algebra
to that of the full theory, the result of which leads to a prediction for the  mass in terms of the scattering length and
contact parameter, which in turn allows us to make a prediction for the quasi-particle lifetime including the normalization.
The width is predicted to scale quadratically with the  ratio of scattering length to the contact parameter.
Note also that the dilaton, because its not derivatively coupled, will only generate the $l=0$ Landau parameter.
Thus we have the additional prediction that the $l=0$  Landau parameter will dominate all other channels. 
We also calculate the compressibility and the spin susceptibilty of the Fermi liquid as a function of the scattering length. 
These  predictions have a limited range of validity. The energy must be small enough that the dilaton exchange can still 
 be treated as a local interaction. This limitation also implies our EFT breaks down when the scattering length, which is inversely proportional to
 the dilaton mass,  becomes  large i.e. in the NFL region. However, our method/prediction is non-perturbative in the
 sense that it is valid up to scattering lengths of order $k_F a \sim 1 $ where perturbative EFT techniques \cite{hammer} fail.
  Figure one summarizes the bounds on the range of validity of the EFT. 
 For typical, model,  values of $\tilde C$ and $T_c$ \cite{contact,Tc}, this gives $k_F a< 5$ for the prediction for the width to be valid.

\begin{figure}
	\centering
	\includegraphics[width=1\linewidth]{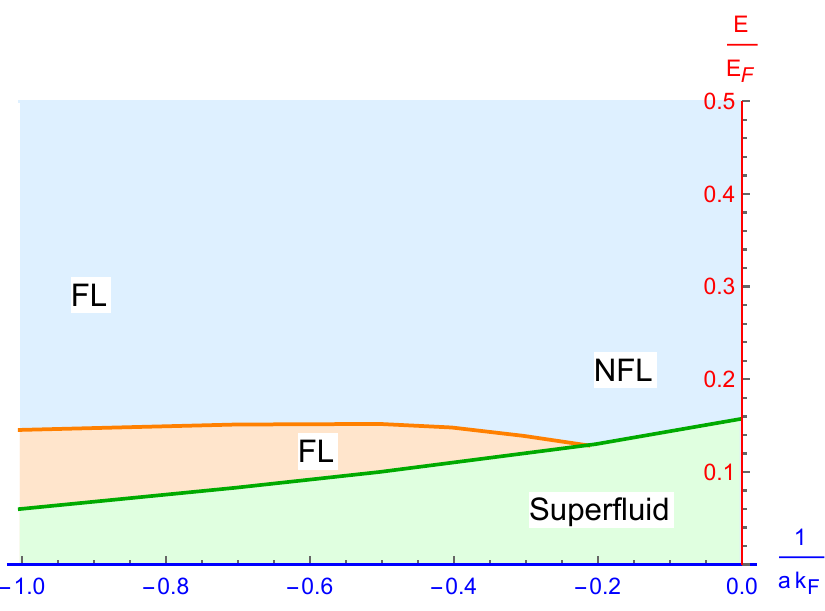}
	\caption{The plot shows the phase diagram as a function of the energy and the scattering length. Our prediction for the self-energy is valid below the dilaton mass curve (orange) and above the $T_c$ curve (green). One obtains a Non-Fermi Liquid (NFL) above the dilaton mass curve for large scattering lengths. As the scattering length decreases, the dilaton starts to decouple and one obtains a conventional Fermi Liquid (FL).}  
	\label{fig:Validity of the EFT}
\end{figure}
\vskip.4in
   \section{Acknowledgments}
We thank  Shauna Kravec, John McGreevy, Cliff Cheung   and Nir Navon for helpful comments on an earlier version of this manuscript.  This work was partially supported by the US Department of Energy under grants   DE- FG02-04ER41338 and FG02-06ER41449.

\end{document}